\documentclass[]{article}
\usepackage{amsmath,graphicx,amssymb,fancyhdr,amsthm,textcomp}  

\newtheorem{thm}{Theorem}[section]

\theoremstyle{definition}

\theoremstyle{remark}

\def\beq{\begin{eqnarray}}
\def\eeq{\end{eqnarray}}
\def\bsp{\begin{split}}
\def\esp{\end{split}}

\newcommand{\la}{{\lambda}}

\newcommand{\Csf}{{\sf C}}
\newcommand{\Rsf}{{\sf R}}
\newcommand{\Tsf}{{\sf T}}

\newcommand{\disc}[3]{{{}_{\sf #1}^{#2}\! D_{#3}}}
\newcommand{\DC}[1]{D_{#1}}
\newcommand{\DT}[1]{\disc{T}{5}{#1}}

\newcommand{\heart}{\ensuremath\heartsuit}

\begin{document}
\title{\Large\textbf{Horizon Detection and Higher Dimensional Black Rings}}
\author{{\large\textbf{A. A.~Coley$^{ \heart}$ and D. D. McNutt$^{\diamondsuit}$}} 
          \vspace{0.3cm} \\ 
                         $^{\heart}$ Department of Mathematics and Statistics,\\ 
                         Dalhousie University, 
                         Halifax, Nova Scotia,\\ 
                         Canada B3H 3J5 
          \vspace{0.3cm}\\
                    $^{\diamondsuit}$ Faculty of Science and Technology,\\ 
                         University of Stavanger, 
                         N-4036 Stavanger, Norway         
		  \vspace{0.3cm}\\
          \texttt{$^{\heart}$ aac@mathstat.dal.ca, $^{\diamondsuit}$ david.d.mcnutt@uis.no} }   
\date{\today}   
\maketitle   
\pagestyle{fancy}   
\fancyhead{} 
\fancyhead[EC]{A. A.~Coley and D. D. McNutt}   
\fancyhead[EL,OR]{\thepage}   
\fancyhead[OC]{}   
\fancyfoot{} 
   
\begin{abstract} 

In this paper we study the stationary horizons of the rotating black ring and the supersymmetric black ring spacetimes in five dimensions. In the case of the rotating black ring we use Weyl aligned null directions to algebraically classify the Weyl tensor, and utilize an adapted Cartan algorithm in order to produce Cartan invariants. For the supersymmetric black ring we employ the discriminant approach and repeat the adapted Cartan algorithm. For both of these metrics we are able to construct Cartan invariants that detect the horizon alone, and which are easier to compute and analyse that scalar polynomial curvature invariants.

\end{abstract}  

\newpage

\section{Introduction}

The study of black holes in general relativity (GR), and the differences between black holes in four dimensions (4D) and higher dimensions, is currently of great interest \cite{HIGHER-D-REVIEW}. At the classical level, gravity in higher dimensions exhibits much richer structure than in 4D; for example, one of most remarkable features of 4D GR is the uniqueness of the Kerr black hole. In contrast, there exist a number of different asymptotically flat, higher-dimensional vacuum black hole solutions \cite{HIGHER-D-REVIEW}. In addition, the horizon structure in higher dimensions is more varied than in 4D. The identification and classification of black hole event horizons is of practical importance in numerical relativity \cite{Thornburg2005}. 

In 4D it has been noted that the invariant $R_{abcd;e} R^{abcd;e}$ vanishes on the horizon for several type {\bf D} solutions \cite{KLA1982}; however, in the case of the Kerr horizon, this invariant detected the stationary limit and not the outer horizon itself. In \cite{AbdelqaderLake2015} the authors examined a collection of invariants from which they determined physical properties of spacetimes around a rotating black hole, including the detection of the horizons. These invariants are obtained from scalar polynomial curvature invariants (SPIs) constructed from the Weyl tensor, its dual and their first covariant derivatives. 

The relationship between these SPIs and the horizons for the Kerr metric was generalized to any stationary black hole \cite{PageShoom2015}, and a general procedure was introduced to determine the location of any stationary horizon for a black hole. While this approach ensures that the event horizon may be detected by a SPI, it does not guarantee that it will not vanish on surfaces outside of the event horizon. 

To pursue an alternative approach to this problem, we consider the Cartan invariants that arise from the Cartan algorithm  \cite{CMcN}. We present an adapted version of the algorithm that will readily produce invariants able to detect the horizon without significant computational effort. These invariants involve the frame components of the Weyl tensor and its first covariant derivatives, and so these expressions will be considerably easier to compute than the related SPIs. They will also be more straightforward to determine if they detect surfaces other than the horizon. We apply the algorithm to detect the stationary horizons of the rotating and supersymmetric black ring (SBR) spacetimes in five dimensions (5D).

\subsection{Algebraic Classification of Spacetimes}

In our approach the Cartan algorithm must at least run through its first iteration before stopping. It is first useful to consider  the algebraic classification of the curvature tensor. The algebraic classification of spacetimes has played a crucial role in the understanding of 4D solutions \cite{kramer}.  

Algebraic classification in 4D can be described in several different ways, using null vectors, 2-spinors, bivectors (or even scalar invariants); each of them can be used to give a different description (or part of) of the 4D algebraic classification scheme. In higher dimensions algebraic classification may be generalized using each of these different methods \cite{class, BIVECTOR}, but each approach leads to a distinct classification in higher dimensions \cite{BIVECTOR,spinorclass}.

When the Ricci curvature tensor vanishes (or is trivial), we need only study a classification of the Weyl curvature tensor.  The most comprehensive and  well-studied approach \cite{class,Alignment,AlignmentReview} classifies null frame components of the Weyl tensor according to their boost weights under local Lorentz boosts by identifying a choice of null frame such that components of higher boost weight vanish. We will refer to the classification of a tensor using its boost weight (b.w.) as the \emph{alignment classification}. We may divide spacetimes into six different primary types:  {\bf G}, {\bf I}, {\bf II}, {\bf III}, {\bf
N}, {\bf O} \cite{class,AlignmentReview,PraPraOrt07,higherghp}.
A spacetime is type {\bf D} if it admits two independent null vector fields with the type {\bf II} property. 

The algebraic types defined by the higher-dimensional alignment classification are quite broad in comparison to the 4D case. The higher-dimensional bivector classification \cite{BIVECTOR} refines  this classification by analysing the bivector map
\begin{equation}\label{map:bivector}
 \Csf: X_{\mu\nu} \mapsto \tfrac{1}{2} C^{\phantom{\mu\nu}\rho\sigma}_{\mu\nu} X_{\rho\sigma},
\end{equation}
which maps the space of spacetime bivectors (2-forms) $X$ to itself.  
In 4D, study of this object reproduces the Petrov classification \cite{kramer}, but in higher dimensions this is not possible. In higher dimensions it was shown that the eigenvector and eigenvalue structure of the bivector map is related to the alignment classification, and allows for a refinement the alignment classification into subtypes \cite{class,BIVECTOR, mahdi}. 

\subsection{Scalar curvature invariants}

A disadvantage of these classification schemes, from a computational perspective, is that a complicated set of equations must be solved in order to find the `preferred' null frame (of
vectors or bivectors) in which to do calculations.  It would be better if there was a constructive way of accessing the invariant classification information.  To do so we introduce the idea of a {\em scalar polynomial curvature invariant of order $k$} (or SPI for short) which is a scalar obtained by contraction from a polynomial in the Riemann tensor and its covariant derivatives up to the order $k$. 

In 4D, SPIs give information on invariant classification \cite{kramer}, and if a spacetime is not locally homogeneous or a member of the degenerate Kundt class (i.e., they are $\mathcal{I}$-non-degenerate, where ${\cal I}$ be the set of all scalar polynomial curvature invariants. ) its SPIs uniquely characterize its geometry  \cite{inv}. It is believed that this is applicable in higher dimensions as well. The black hole spacetimes do not belong to the degenerate Kundt class and so they are ${\cal I}$-non-degenerate.

Similar algebraic techniques to those in \cite{kramer} were developed in higher dimensions using `discriminants' \cite{DISCRIM}. It was shown that an explicit algorithm can be used to completely determine the eigenvalue structure of the curvature operator, up to degeneracies, in terms of a set of discriminants \cite{DISCRIM}. These conditions on the discriminants can be expressed in terms of the SPIs, and these techniques can be used to study the necessary conditions (syzygies) in arbitrary dimensions for the Weyl curvature operator (and hence the higher dimensional Weyl tensor) to be of algebraic type {\bf II}, {\bf D} or more special \cite{CHDG}.

\subsection{Algebraic Classification of Curvature Operators Using Discriminants} \label{sec:disc}

In this subsection we briefly review the scalar invariant approach to the algebraic classification of spacetimes introduced in \cite{DISCRIM}, which provides a link between SPIs and bivector operators \cite{BIVECTOR} and the alignment classification \cite{class}. As an application of this technique, we can construct invariants to identify the horizons of black hole spacetimes \cite{AbdelqaderLake2015,PageShoom2015}. All black
holes are of type {\bf II} or more special on the horizon 
\cite{Lewandowski:2004sh} and, in general, the relevant discriminants for the black holes studied are non-vanishing outside the horizon, and hence the spacetime must be of type {\bf G} or {\bf I} in the exterior region. 

For any curvature tensor, there are several operators describing  automorphisms of finite-dimensional vector spaces that can be defined.  The most well-known example of such a map is the
bivector map (\ref{map:bivector}) obtained from the Weyl tensor which acts on the $D(D-1)/2$-dimensional vector space of bivectors. If the Weyl tensor is of a particular algebraic type in the alignment classification, the associated operator $\Csf$ will have a restricted eigenvector structure \cite{BIVECTOR,OP}. The Ricci operator could be used also; however, for many black hole spacetimes this map is trivial. 

This technique can be extended to include any $\Rsf \in {\cal R}$, where ${\cal R}$ denotes the set of all curvature operators constructed from the curvature tensors, their covariant derivatives, and their polynomial invariants. A particularly useful example arises from the tensor
\begin{equation}
  T^{\alpha}_{~\beta} \equiv C^{\alpha\mu\nu\rho}C_{\beta\mu\nu\rho},
\end{equation}
\noindent which can be used to construct an operator 
\begin{equation}  \label{eqn:Tdef}
  \Tsf: X^\alpha \mapsto T^{\alpha}_{~\beta} X^\beta - I_2 \delta_\alpha^\beta,~~~ I_2 = C^{\alpha\mu\nu\rho}C_{\alpha\mu\nu\rho},  
\end{equation}
acting on the $D$-dimensional tangent space of the spacetime. 

\subsubsection{Discriminant analysis}
For any relevant curvature operator, ${\sf R}$, we can examine the eigenvalues of the operator to obtain necessary conditions on various alignment types of the spacetime.  The algebraic types {\bf II} or {\bf D}  will impose restrictions on the eigenvalues of the resulting operator \cite{DISCRIM}.  To see this, we examine the characteristic equation
\begin{equation} 
  f(\la) \equiv \det (\Rsf - \lambda {\sf 1} ) = 0; 
\end{equation} 
this is a polynomial equation in $\lambda$ and the eigenvalues will be the roots of this equation.  Each coefficient of this polynomial can be expressed in terms of the invariants of $\Rsf$.  
Therefore, we can give conditions on the eigenvalue structure expressed in terms of the polynomial curvature invariants of ${\sf R}$.

\subsubsection{Necessary conditions for Weyl type {\bf II}/{\bf D}} 

Requiring that the Weyl tensor is of type {\bf II} (or more
special) causes the eigenvalues of the corresponding bivector operator $\Csf$ in \eqref{map:bivector} to be of a special form. Noting that the conditions for type {\bf D} provide a similar set of invariants as that of type {\bf II}, we make the assumption the spacetime is type {\bf D} in the following.  In \cite{BIVECTOR,DISCRIM}, 
it was shown that in the type {\bf D} case the Weyl bivector operator for a spacetime of type {\bf II} or more special has $(D-2)$ eigenvalues of multiplicity 2 at the very least.

In 5D, the bivector operator acts on a vector space that is $10$-dimensional and if the spacetime is type {\bf II} then the bivector operator has 3 eigenvalues of (at least) multiplicity 2.  Hence the eigenvalue (Segre) type of the matrix is $\{(11)(11)(11)1111\}$ (or more special, e.g.\ $\{(1111)(11)1111\}$). Applying the discriminant analysis, it was 
shown in \cite{DISCRIM} that eigenvalue types consistent with at least 3 matching  pairs of eigenvalues can only occur if the following three discriminants vanish:  
  \begin{equation}
    \DC{8} = 0, \qquad
    \DC{9} = 0, \qquad 
    \DC{10} = 0.
  \end{equation}  
These three discriminants (which are explicitly defined in \cite{DISCRIM}) are syzygies of order 90, 72 and 56, respectively.  While it is conceivable that these can be computed  and written out in full in terms of the Weyl tensor in full generality, in practice this is not feasible and it is far more useful to apply the algorithm used to construct the discriminants for a particular choice of metric.

As an illustration, let us consider the operator $\Tsf$ defined in (\ref{eqn:Tdef}); the following conditions on the discriminants hold:
  \begin{equation}
    \DT{5}=0, \qquad
    \DT{4} \geq 0, \qquad
    \DT{3} \geq 0, \qquad 
    \DT{2} \geq 0.
  \end{equation} 
Note that $\DT{5}=0$ is a 40th order syzygy in the Weyl tensor (a 20th order syzygy in the square of the Weyl tensor), and hence is likely to be easier to calculate explicitly than the syzygies resulting from the bivector operator.  

In five dimensions (5D), the discriminant approach was applied to the 5D supersymmetric rotating black  holes (SBR)  \cite{SBR} (that includes the extremal charged rotating BMPV black hole of \cite{BMPV}). It was shown that the vanishing of the discriminants defined in equation \eqref{discInv} is consistent with the SBR spacetime being of type {\bf II/D} on the horizon \cite{CHDG}. As long as the discriminant, $\DT{5}$, defined in \eqref{discInv}  is non-zero, the Weyl tensor is of type {\bf I/G}. It is feasible that we can use these SPIs to identify the horizons of higher dimensional black holes, as they are  necessarily of Weyl type {\bf II/D}.

%
%
%

\newpage


\subsection{A brief review of the Cartan algorithm in 5D}

While the algorithm has been implemented in 4D \cite{kramer}, we may also apply the Cartan algorithm to determine a set of Cartan invariants in higher dimensions. The aim of the Cartan algorithm is to reduce the frame bundle to the smallest possible dimension at each step by casting the curvature and its covariant derivatives into some preferred form and only permitting those frame changes which preserve the form of the curvature tensor and its covariant derivatives. The standard algorithmic procedure for determining equivalence is given in  \cite{CMcN} but is restated here for convenience: 

\begin{enumerate}
\item Set the order of differentiation $q$ to 0.
\item Calculate the derivatives of the Riemann tensor up to the $q^{th}$ order.
\item Determine the preferred form of the Riemann tensor and its covariant derivatives.
\item Fix the frame as much as possible while preserving the preferred form, and note the residual frame freedom (the group of allowed transformations is the {\it linear isotropy group $\hat{I}_q$}). The dimension of $\hat{I}_q$ is the dimension of the remaining {\it vertical} freedom of the frame bundle.
\item Find the number $t_q$ of independent functions of spacetime position in the components of the Riemann tensor and its covariant derivatives. This tells us the remaining {\it horizontal} freedom.
\item If the isotropy group and number of independent functions are the same as in the previous step, let $p+1=q$, and the algorithm terminates; if they differ (or if $q=0$), increase $q$ by 1 and go to step 2. 
\end{enumerate}

\noindent When applying this algorithm, we must choose a preferred form for the curvature tensor by fixing the frame. This requires the study of the transformation properties of the Riemann tensor and its covariant derivatives under Lorentz transformations (spins, boosts and null rotations) \cite{CMcN}. The resulting non-zero components of the curvature tensor and its covariant derivatives relative to the frame at the end of the algorithm are called {\it Cartan invariants}. 

The $D$-dimensional spacetime is then characterized by the form of the curvature tensor and its covariant derivatives, the successive isotropy groups and the independent function counts.  As there are $t_p$ essential spacetime coordinates, the remaining $D-t_p$ are ignorable, and so the dimension of the isotropy group of the spacetime will be $s=\dim(\hat{I}_p)$, and the isometry group has dimension $r=s+D-t_p$.

\newpage 

\section{Locating Stationary Horizons with Invariants}

%
%

The procedure for constructing invariants that detect the event horizon of stationary black holes \cite{PageShoom2015} relies on the fact that the squared norm of the wedge product of $n$ gradients of functionally independent locally smooth curvature invariants will always vanish on the horizon of any stationary black hole. Here, $n$ is the local cohomogeneity of the metric, which is the codimension of the maximal dimensional orbits of the isometry group of the local metric.  The result may be summarized in the  following theorem \cite{PageShoom2015}:

\begin{thm} \label{PSthrm}
For a spacetime of local cohomogeneity $n$ that contains a stationary horizon, $\mathcal{H}$ (a null hypersurface that is orthogonal 
to a Killing vector field that is null there and hence lies within the hypersurface and is its null generator), 
and which has $n$ scalar polynomial curvature invariants $S^{(i)}$ whose gradients are well-defined there, the $n$-form wedge product 
\beq W = dS^{(1)} \wedge ... \wedge dS^{(n)} \nonumber \eeq
\noindent has zero squared norm on the horizon, 

\beq ||W||^2 = \frac{1}{n!} \delta^{\alpha_1,...,\alpha_n}_{\beta_1,...,\beta_n} g^{\beta_1 \gamma_1} ... g^{\beta_n \gamma_n} \times S^{(1)}_{;\alpha_1}...S^{(n)}_{;\alpha_n} S^{(1)}_{;\gamma_1}...S^{(n)}_{;\gamma_n} \biggr|_{\mathcal{H}} = 0. \nonumber \eeq
\noindent Where the permutation tensor $\delta^{\alpha_1,...,\alpha_n}_{\beta_1,...,\beta_n}$ is $+1$ or $-1$ if $\alpha_1,...,\alpha_n$ is an even or odd permutation of $\beta_1,...,\beta_n$ respectively, and is zero otherwise. 
\end{thm} 

The above theorem can be generalized to the Cartan invariants arising from the equivalence algorithm by replacing the $n$ SPIs $S^{(i)}$ by $n$ functionally independent Cartan invariants. This is illustrated in the resource article \cite{GANG} where several exact black hole solutions are studied using both scalar polynomial curvature invariants and Cartan invariants. 

The examples presented in \cite{GANG} are fairly simple exact solutions, and so the associated SPIs and Cartan invariants for the metrics are compact expressions that are easily studied. In practice, for more complicated metrics, the SPIs are considerably larger which complicates their analysis. In general, the Cartan invariants will be much simpler to compute than the related SPIs and hence are a helpful alternative to SPIs.

\subsection{Algorithm for Generating Invariants for Event Horizon Detection}

The algorithm we propose for generating invariants for horizon detection will not necessarily be an optimal application of the Cartan algorithm. Typically the algorithm is intended for comparing two spacetimes that may be equivalent. If the two sequences $\{\hat{I}_q\}$ and $\{ t_q\}$ match for both spacetimes we would want to compare the Cartan invariants. This is best accomplished by choosing a frame in which the least number of non-zero components appear at each iteration. This allows for a coordinate independent way to specify the frame; however, choosing such a frame is not always simple \cite{5Dclass}.

Instead, we will choose a frame most convenient for computation of the invariants relative to a given coordinate system. This is permissible because the Cartan algorithm does not specify a canonical form for the Riemann tensor and its covariant derivatives, and our goals are not solely focused on the classification of spacetimes; we just need to fully fix the frame to obtain invariants.  

The algorithm consist of four steps:
\begin{enumerate}
\item Choose a coframe that diagonalizes the metric.
\item Compute the $q=0$ iteration of the Cartan algorithm, and choose the frame such that the components of $R_{abcd}$ are easily computed \footnote{Frequently fixing the frame with the identity frame transformations is best; for some examples, like 5D Kerr-AdS in Kerr-Schild form, choosing the frame for which the metric is type {\bf D} is the best choice from a computational perspective \cite{CMcN}.}.
\item Compute the $q=1$ iteration of the Cartan algorithm, and again choose the frame such the the components of $R_{abcd;e}$ are easily computed.
\item Choose the smallest components of $R_{abcd;e}$ which vanishes on the horizon. The existence of such invariants is ensured by Theorem \ref{PSthrm}.
\end{enumerate}

\newpage
\section{Five Dimensional Rotating Black Ring} \label{sec:RBR}

The rotating black ring solution\footnote{This solution has an interesting horizon topology $ S^1 \times S^2 $, which cannot occur in 4D.} was found in \cite{RBR}; however, we will use the coordinates $(t,x,y,\phi, \psi)$ introduced in \cite{EE:2003}:

\beq ds^2 &=& - \frac{F(x)}{F(y)} ( dt + R\sqrt{\lambda \nu} (1+y)d\psi)^2 \nonumber \\
&&+ \frac{R^2}{(x-y)^2} \left[ - F(x) \left( G(y) d\psi^2 + \frac{F(y)}{G(y)} dy^2 \right) + F(y)^2 \left( \frac{dx^2}{G(x)} + \frac{G(x)}{F(x)} d\phi^2 \right) \right] \nonumber \eeq
\noindent where
\beq F(\zeta) = 1 -\lambda \zeta,~~~~~~~ G(\zeta) = (1-\zeta^2)(1-\nu \zeta). \nonumber \eeq

Looking at the eigenvalues of the metric, the only regions for $(x,y)$ with signature (1,4) are the following:
\begin{itemize}
		\item Region $\mathcal{A}_1:~(-1,1) \times (-\infty, -1)$ is asymptotically flat and static. This represents the outer part of the black ring solution, this regions can be smoothly connected with $\mathcal{A}_2$ by identifying $y = -\infty$ with $y = \infty$.
		\item Region $\mathcal{A}_2:~(-1,1 ) \times (1/\nu, \infty)$ represents an ergosphere with a limiting surface of stationarity located at $y = \infty$ and a horizon at $y = 1/\nu$.
		\item Region $\mathcal{A}_3:~(-1,1) \times (1/\lambda, 1/\nu)$ is non-stationary and represents the region below the horizon, the curvature singularity occurs at $y = 1/ \lambda$.
		\item Region $\mathcal{B}:~(-1,1) \times (1,1/\lambda)$ represents the region around a spinning singularity. 
		\item Regions $\mathcal{C}_1:~(1/\lambda, 1/\nu) \times (-1,1)$ and $\mathcal{C}_2:~(1/\nu, \infty) \times (-1, 1)$ are not asymptotically flat, and their physical interpretation is unknown.        
\end{itemize}        

\subsection{Analysis of Weyl Aligned Null Directions}

The rotating black ring is a vacuum solution, and so the Ricci tensor vanishes. We can use the b.w. decomposition \cite{BIVECTOR, CH2011,CHOL2012} to classify the Weyl tensor. For the Weyl tensor, the indicial symmetries imply that all components of b.w. $\pm 4$ or $\pm 3$ are zero, and the remaining components of b.w. satisfy several algebraic relations \cite{CHOL2012}. In 5D this classification can be made finer by considering the spin group which is isomorphic to $O(3)$ and acts on the null frame according to \beq \ell' = \ell ,~~n' = n,~~m^{i'}
= m^j G_j^{i} \nonumber \eeq \noindent The details of  this classification are expanded upon in \cite{CHOL2012}. This classification can also be applied to the covariant derivative of the Wey tensor, and so it will allow us to determine the isotropy group at the zeroth and first iteration.

To determine the alignment classification for the rotating black ring, we exploit the Weyl Aligned Null Directions (WANDs). In \cite{brwands}, the higher dimensional Bel-Debever criteria \cite{bdcond} was used to show that the rotating black ring is generally of type ${\bf I}_i$, so that the components of b.w. $+2$ and $-2$ may be set to zero by the choice of an appropriate frame, leaving the following algebraically independent components of Weyl:
\beq C_{0101},~ C_{0ijk},~ C_{1ijk},~ C_{0i1j}, \text{ and } C_{01ij},~ i,j,k \in [2, 4]. \label{typeIi} \eeq 
\noindent On the horizon, the rotating black ring is of type {\bf II}; this may be seen by noting that the Weyl components
\beq C_{0ijk}, \text{ and } C_{1ijk}  \nonumber \eeq
\noindent vanish relative to the above frame where the Weyl tensor is of type ${\bf I}_i$. 

In particular, the equations used to determine the WANDs for the stationary region of the rotating ring were presented in \cite{brwands}. This can be confirmed by showing that the discriminants discussed in section \ref{sec:disc} only vanish on the horizon \cite{CH}; i.e., 
\beq
  \DC{4} = \DC{5} = \DC{6} = \DC{7} = \DC{8} = \DC{9} = \DC{10} = 0. \nonumber \eeq

\subsection{Event Horizon Detection}

For the region $\mathcal{A}_2$, we choose the following orthonormal tetrad: 
\small
\beq & t^0 =  \sqrt{-\frac{F(x)}{F(y)}} ( dt + R\sqrt{\lambda \nu} (1+y)d\psi),~~ t^1 = \frac{R}{(x-y)} \sqrt{F(x) G(y)} d\psi & \nonumber \\
& t^2 = \frac{R}{(x-y)} \sqrt{ \frac{F(x) F(y)}{G(y)}} dy,~~
t^3 = \frac{R F(y)}{(x-y)}  \frac{dx}{\sqrt{G(x)}},~~t^4 = \frac{R}{(x-y)} \sqrt{\frac{G(x)}{F(x)}} d\phi &  \nonumber \eeq
\normalsize
\noindent With this orthonormal coframe, we construct the null coframe:
\small
\beq \ell = \frac{t^0 + t^1}{\sqrt{2}},~~n = \frac{- t^0 + t^1}{\sqrt{2}}, m^2 = t^2,~~m^3 = t^3,~~m^4 = t^4. \nonumber \eeq
\normalsize
\noindent While the components of the Weyl tensor are too large to display compactly, we may discuss their structure in a symbolic fashion. Relative to this frame, the algebraically independent non-zero components of the Weyl tensor are \cite{CHOL2012}:
\small 
\beq & C_{0101}, ~C_{0123},& \nonumber \\
& C_{0i1j} = \left[ \begin{array}{ccc} C_{0212} & C_{0213} & 0 \\ 
-C_{0213} & C_{0212} & 0 \\ 0 & 0 & C_{0414} \end{array}\right],  C_{AiAj} = \left[ \begin{array}{ccc} C_{A2A2} & C_{A2A3} & 0 \\ 
C_{A2A3} & C_{A3A3} & 0 \\ 0 & 0 & C_{A4A4} \end{array}\right],& \nonumber\eeq
\normalsize
\noindent where $i,j \in [2,4]$ and $A = 0,1$. Notice that this frame is not constructed from WANDs as the Weyl tensor has non-zero b.w. $+2$ and $-2$ components. Furthermore, the submatrix, $C_{Ai'Ai'},~i'=2,3$ may be diagonalized with equal eigenvalues. 

While the Weyl tensor is of type {\bf II} on the horizon, knowing that the Weyl tensor is of type ${\bf I}_i$ elsewhere, allows us to determine the isotropy group at zeroth order. At zeroth order all frame transformations except a spatial rotation about $m^4$ affect the form of the metric \cite{CMcN}. Fixing the frame using the identity frame transformations ensures the components of the Weyl tensor will be Cartan invariants. 

Computing the components of the first covariant derivative of the Weyl tensor, any rotation about $m^4$ will affect the form of $C_{0101;i},~i\in [2,3]$. We may fix the remaining spatial rotation to identity. Examining the components of $C_{abcd;e}$, we find two components that vanish on the event horizon alone, due to the constraints on $(x,y)$ in region $\mathcal{A}_2$:

\beq & C_{0104;2} = \frac {-3/4\, \left( \lambda-\nu \right)  \left( x-y \right) ^{3} \left( \lambda-1
 \right)  \left( \lambda+1 \right) \sqrt { \left( y-1 \right)  \left( 1+y
 \right) }\sqrt {\nu y-1}}{ \left( 1 - \lambda x \right) ^{3/2}{R}^{3} \left( \lambda y-1
 \right) ^{9/2}} & \label{RBRinv} \\ 
 & C_{0134;4} = \frac12 \frac { \left( x-y \right) ^{2}\sqrt { \left( x-1 \right) 
 \left( x+1 \right) }\sqrt {\nu x-1}\sqrt {\lambda}\sqrt {\nu}\sqrt { \left( y-1
 \right)  \left( 1+y \right) }\sqrt {\nu y-1}}{ \left( \lambda y-1 \right) ^{7/2
}{R}^{3}} &\nonumber \eeq

When $y = \frac{1}{\nu}$ these invariants vanish on the horizon. It is worthwhile to examine the other potential roots of this equation: 
\begin{itemize}
\item If $\lambda = \nu$, $C_{0104;2}$ would vanish. However, $\lambda \neq \nu$ as $\nu < \lambda$ in order to a horizon topology $S^2 \times S^1$. 
\item A root exists if $x =y$; however, the line $y=x$ does not lie in any region where the metric has signature $(-++++)$ and is non-flat.
\item A root exists if $\lambda =1$; however, setting $\lambda = 1$ gives the singly spinning rotating Myers Perry solution. The structure of the Cartan invariants will be different in this case \cite{CMcN}. 
\item A root exists if $y = 1$; however, $y$ is restricted to the open interval $(-1,1)$. 
\end{itemize} Notice that by taking the limit as $y \to \pm \infty$ of the ratio $\frac{C_{0134;4}}{C_{0104;2}} \to 0 $, we detect the stationary limit. 

In principle, the Cartan algorithm must be completed to relate  these quantities to the invariants from the Cartan algorithm carried out in an optimal fashion. The frame producing a minimal set of Cartan invariants would be ideal in which to compute discriminants, as there would be the least number of non-zero components at each iteration.

\subsection{Singly Rotating Myers-Perry Black hole}

We note that that the singly rotating Myers-Perry black hole solution is a special case of the rotating black ring, when the parameter $\lambda$ is set equal to unity. In this case the zeroth order invariants cannot detect the horizon as this solution is of type {\bf II/D}; however, the invariants in \eqref{RBRinv} still detect the horizon. However, the SPI approach is just as straightforward to implement, and so it does not provide a suitable example to illustrate the utility of the modified algorithm.

\newpage
\section{Supersymmetric Black Ring}

The algorithm we have developed may be applied to non-vacuum spacetimes as well. We now consider an interesting class of non-vacuum metrics, corresponding to the supersymmetric, rotating black holes in 5D. The simplest example of such a spacetime is the extremal, charged, rotating BMPV black hole \cite{BMPV} in 5D minimal supergravity, which has a horizon of spherical topology.

The BMPV metric is generally of Weyl type {\bf I}$_i$ \cite{5Dclass}; therefore, there is a frame where the Weyl tensor will have non-zero components as in equation \eqref{typeIi}. On the horizon the metric is of type {\bf II}, which may be seen by choosing a particular frame where the components $C_{0ijk}, \text{ and } C_{1ijk},~i,j,k\in [2,4]$ vanish on the horizon. As a generalization of the BMPV metric,  a supersymmetric black ring solution of the same supergravity theory was presented in \cite{SBR}, with a black hole horizon of topology $S^1 \times S^2$.  The BMPV metric can be obtained as a particular limit of the supersymmetric black ring family.  

As the BMPV metric is a particular case of the supersymmetric black ring (having a horizon topology $S^{1}\times S^{2}$), one would expect that the supersymmetric black ring is not of Weyl type {\bf II} or more special. This was proven in \cite{{CHDG}} where the discriminant method was used to determine that it is indeed of type {\bf I/G}. However, unlike the rotating black ring, the WANDs have yet to be determined explicitly for this solution. 

The line element of the supersymmetric black ring is
\begin{equation}
  ds^2= -f(x,y)^2 (dt+\omega)^2 + f(x,y)^{-1} ds^{2}(\mathbf{R}^{4}),
\end{equation}
where
\begin{eqnarray}
  f(x,y)^{-1}  &\equiv & 1+\tfrac{Q-q^2}{2R^2}(x-y)-\tfrac{q^2}{4R^2}(x^2-y^2),\\ 
  \omega       &=      & \omega_{\phi}d\phi+\omega_{\psi}d\psi , \label{deff} \\
  \omega_{\phi}&=      & -\tfrac{q}{8R^2}(1-x^2)[3Q-q^{2}(3+x+y)], \\
  \omega_{\psi}&=      & \tfrac{3}{2}q(1+y)+\tfrac{q}{8R^2}(1-y^2)[3Q-q^{2}(3+x+y)],
\end{eqnarray}
and the four dimensional flat space metric is written as
\begin{equation}
  ds^2(\mathbf{R}^{4}) = \frac{R^2}{(x-y)^2}\left[\frac{dy^2}{y^2-1}+(y^2-1)d\psi^2
                                                  +\frac{dx^2}{1-x^2}+(1-x^2)d\phi^2\right].
\end{equation}
Admissible coordinate values are $-1\leq x \leq 1$, $-\infty < y \leq -1$ and $\phi$, $\psi$ are $2\pi$-periodic, and the parameters $q$ and $Q$ satisfy $q>0$ and $Q \geq q^2$ (which implies that $f(x,y)>0$). The black ring horizon lies at $y=-\infty$ in these coordinates.  The parameters $R$, $Q$ and $q$ must satisfy the following inequality in order to ensure closed timelike curves do not exist in the solution \cite{EE:2003}: 

\beq R < \frac{Q-q^2}{2q} \label{SBRcon} \eeq


\subsection{Discriminant Analysis} \label{subsec:SBRda}

The discriminant analysis was presented for a particular set of parameter values in \cite{CHDG}. The parameter $R$
fixes the length-scale, and hence was set to $R=1$ without loss of generality.  The discriminants were computed for various values of the parameters $q$ and $Q$; in particular, the case where $q=1/2$, $Q=9/4$ was presented in \cite{CHDG}. 

For the Weyl bivector operator $\Csf$, the discriminants can be evaluated on the horizon,  $y=-\infty$, and we obtain:
\begin{equation}
  \DC{1} = 10, \ 
  \DC{2} = 180, \
  \DC{3} = 4608; \ \
  \DC{4} = \DC{5} = \DC{6} = \DC{7} = \DC{8} = \DC{9} = \DC{10} = 0.
\end{equation}
This is consistent with the black ring being type {\bf II} or {\bf D} on the horizon. The eigenvalue type of the Weyl bivector operator is of the form $\{1(111)(111111)\}$ on the horizon, while the operator $\Tsf$ is of eigenvalue type $\{(11)(111)\}$ there.

Off the horizon, the discriminants are considerably more complicated.  Computing the operator $\Tsf$ indicates that the discriminant $\DT{5}$ takes the form:
\begin{equation}
  \DT{5} = F(x,y) \frac{(y-1)^2 (y+1)^2 (x-1)^2 (x+1)^2 f(x,y)^{154}}{(x-y)^{70}}, \label{discInv}
\end{equation}
where $F(x,y)$ is some (known) polynomial in $x$ and $y$ and $f$ is the metric function defined in (\ref{deff}).  Wherever $\DT{5}$ is non-zero, the Weyl tensor must be of type {\bf I} or {\bf G}.

Checking that $F(x,y) \neq 0$ is relatively difficult. It was verified that this is true for the following cases (among others): 
(i) $x=0$, $y$ general. 
(ii) $x=\pm 1/2$, $y$ general. 
(iii) $y=-2$, $x$ general.
This shows that the metric is of type {\bf I/G} except possibly for a few special values of
$(x,y)$. 
For the supersymmetric black ring, some of the special cases are as follows:  $x=y$: which represents a curvature singularity, at x=y=-1: (for the coordinate ranges used here) corresponding to asymptotic infinity,  $x=\pm 1$: the plane of the ring
(the axis of $\phi$ rotation), and $y=-1$: the axis of $\psi$-rotation.

\subsection{Event Horizon Detection}

We choose the following orthonormal tetrad, 

\beq & t^0 = f(x,y)^2 (dt+\omega),~~ t^1 = \sqrt{\frac{f(x,y)}{y^2-1}} \frac{R dy}{x-y} &  \nonumber \\
& t^2 = \frac{\sqrt{f(x,y)(y^2-1)}R}{x-y} d \psi ,~~ t^3 = \sqrt{\frac{f(x,y)}{(1-x^2)}}  \frac{R dx}{x-y},~~  t^4 = \frac{\sqrt{f(x,y)(y^2-1)}R}{(x-y)}  d\phi &  \nonumber \eeq

\noindent With this orthonormal coframe, we construct the null coframe:
\beq \ell = \frac{t^0 + t^1}{\sqrt{2}},~~n = \frac{ t^0 - t^1}{\sqrt{2}}, m^2 = t^2,~~m^3 = t^3,~~m^4 = t^4. \nonumber \eeq

\noindent The Weyl tensor will be of type ${\bf I/G}$ except on the horizon where it will be type {\bf II}; hence, off the horizon, the isotropy group at zeroth order is at most a spatial rotation \cite{BIVECTOR}. Fixing the frame using the identity frame transformations ensures the components of the Weyl tensor will be Cartan invariants. 

Calculating $C_{abcd;e}$, and fixing any potential spatial rotation to identity ensures the resulting components are Cartan invariants. We find that the smallest first order Cartan invariant that vanishes on the event horizon is:

\beq C_{2324;2} = \frac {4/3\, \left( x-y \right) ^{4}{q}^{4} P(x,y) \sqrt { \left( x+1
 \right)  \left( 1-x \right) }}{ \left( 2\,xQ-2\,{q}^{2}x-{x}^{2}{q}^{
2}+4\,{R}^{2}-2\,yQ+2\,{q}^{2}y+{y}^{2}{q}^{2} \right) ^{11/2}}
\nonumber \eeq

\noindent where $P(x,y)$ is a degree four polynomial in $x$ and $y$:
\small
\beq && 36\,y{Q}^{2}x+42
\,{y}^{3}{q}^{4}x-264\,{x}^{3}{R}^{2}{q}^{2}+72\,{x}^{2}Q{q}^{2}-156\,
{x}^{4}{R}^{2}{q}^{2}-96\,{x}^{3}{R}^{2}y{q}^{2} \nonumber \\
&&-24\,{q}^{2}{R}^{2}y{x}^{2} -36\,{q}^{2}{R}^{2}{y}^{2}{x}^{2}+24\,y{x}^{2}{R}^{2}Q+102\,{q}^{
2}Q{y}^{2}{x}^{3}-18\,{q}^{2}Q{y}^{3}{x}^{2} \nonumber \\
&& -24\,{q}^{2}Q{y}^{2}{x}^{2
}-78\,{q}^{2}{x}^{4}Qy+120\,{q}^{2}Q{x}^{3}y+96\,{R}^{2}y{q}^{2}x-92\,
{y}^{2}Q{q}^{2}x\nonumber \\
&&-72\,yQ{q}^{2}x+76\,yQ{x}^{2}{q}^{2}+53\,{x}^{4}{q}^{4
}-8\,{q}^{4}{y}^{3}-13\,{y}^{4}{q}^{4}-76\,{x}^{2}y{q}^{4}\nonumber \\
&&+244\,{R}^{2
}{x}^{2}{q}^{2}-12\,{q}^{2}xQ-60\,y{x}^{3}{Q}^{2}+12\,{x}^{2}{Q}^{2}{y
}^{2}-96\,{q}^{2}{x}^{4}Q-102\,{x}^{3}y{q}^{4}\nonumber \\
&&+20\,{y}^{2}{q}^{4}{x}^{
2}-240\,{R}^{2}xQ+8\,{x}^{3}Q{q}^{2}-48\,{R}^{2}yQ+12\,{q}^{2}yQ+42\,{
q}^{4}{x}^{5}y+36\,{q}^{4}yx\nonumber \\
&&+92\,{q}^{4}{y}^{2}x+15\,{q}^{4}{y}^{4}{x}
^{2}+240\,{R}^{2}{q}^{2}x+78\,{q}^{4}{x}^{4}y+18\,{q}^{4}{y}^{3}{x}^{2
}+48\,{R}^{2}{q}^{2}y\nonumber \\
&&+44\,{R}^{2}{y}^{2}{q}^{2}+8\,{y}^{3}Q{q}^{2}-42
\,{q}^{4}{y}^{3}{x}^{3}+96\,{R}^{4}-6\,{q}^{4}{x}^{4}{y}^{2}-102\,{q}^
{4}{y}^{2}{x}^{3}+12\,{q}^{4}x\nonumber \\
&&-96\,{R}^{2}{q}^{2}-6\,{q}^{4}{y}^{2}-30
\,{q}^{4}{x}^{2}-12\,{q}^{4}y-8\,{q}^{4}{x}^{3}-6\,{q}^{2}{x}^{5}Q+264
\,{x}^{3}{R}^{2}Q+6\,{q}^{4}{x}^{5}\nonumber \\
&&-9\,{q}^{4}{x}^{6}+48\,{x}^{4}{Q}^{
2}-36\,{x}^{2}{Q}^{2}-96\,{R}^{4}{x}^{2}  \nonumber \eeq
\normalsize
Taking the limit $y \to - \infty$ this vanishes, and hence detects the horizon. In principle, the Cartan algorithm must be completed to relate  these quantities to the Cartan invariants produced by the Cartan algorithm carried out in an optimal fashion. 

To study the zeros of the $P(x,y)$, we consider the case where $R=1$ and $Q = 2q^2+1$ in order to satisfy \eqref{SBRcon}; this produces a degree four polynomial in terms of $q$. This polynomial has one real-valued root for $q$ which is  greater than zero for any value of $x\in [-1,1]$ and $y \in (-\infty,-1]$. Fixing $q$ to be some constant value, this shows that for each positive $q$ there will be curves in the $x-y$ plane where $P(x,y;q) = 0$. Note that these curves are degenerate, as they approach $x=\pm1$ as $y \to -\infty$ which corresponds to the axis of rotation about $\phi$. 

It is still possible to construct an invariant that vanishes on the horizon alone. Consider the component $C_{0102;4}$, which will have a similar structure to $C_{2324;2}$ albeit with a different corresponding degree four polynomial. Repeating the above analysis, we may show that the zeros of this polynomial differ from that of $P(x,y)$ and so the invariant 
\beq [C_{2324;2}]^2 + [C_{0102;4}]^2  \nonumber \eeq
\noindent will only vanish on the horizon for all values of $q$.

\section{Discussion}

Employing SPIs for horizon detection can be difficult to implement in practice. We have introduced an algorithm adapted from the Cartan algorithm intended for the calculation of invariants that detect the horizon of stationary black holes in five dimensions. To demonstrate the applicability of the algorithm, we have applied the algorithm to two physically interesting solutions: the rotating black ring and the supersymmetric black ring. These solutions are sufficiently complicated to motivate this approach. We also note that the Myers-Perry black hole, which is of type {\bf II}, is included as a special case of the rotating black ring solution.

Using the discriminant analysis to determine the algebraic type at the zeroth iteration of the algorithm, the adapted algorithm allows us to construct scalar invariants from fixing the frame in some trivial manner and computing the next iteration of the Cartan algorithm. For both spacetimes, the resulting invariants are considerably less complicated than any corresponding SPIs, and they vanish on the stationary horizons of the black rings. 

The frame approach used here is not quite the Cartan algorithm. To apply this to the equivalence problem, we would have to fix the frame at each iteration in a clear coordinate independent manner, and record the dimension of the isotropy group, and number of functionally independent invariants at each order until the algorithm terminates. In contrast the modified algorithm terminates at first iteration, and the choice of frame is dependent on the coordinate system, which is all that is necessary for the purpose of horizon detection.

With the conditions on frame fixing relaxed for the Cartan algorithm, the Cartan invariants are much easier to compute than the related SPIs, and this makes the identification of the horizon computationally easier. In the case of the rotating black ring spacetime, we are able to construct a Cartan invariant that will vanish on the horizon alone. The supersymmetric black ring provides an example where a chosen Cartan invariant also vanishes on surfaces outside of the event horizon; however, due to the lower degree polynomials involved, we are able to build an invariant that vanishes on the horizon alone by taking the sum of two horizon detecting Cartan invariants.

All black hole spacetimes are of type {\bf II/D} on the horizon.  In general a (stationary) black hole spacetime is of type {\bf I} or {\bf G} and so the zeroth order type {\bf II/D} discriminant SPIs (or the corresponding Cartan invariants) vanish on the horizon and are non-vanishing elsewhere (and hence signal the horizon).  The Page-Shoom results indicate that a first order SPI (or the corresponding results using first order Cartan invariants studied here) locates the horizon.  In the type {\bf II/D} spacetimes considered in \cite{GANG} it was necessary to identify the horizon using a first order invariant.  However, in the black ring spacetimes considered in this paper, which are both generally of type $ {\bf I}_i $, not only do the type {\bf II/D} zeroth order invariants vanish on the horizon, but the first order invariants do as well (and, in addition, can be used to
uniquely identify the horizon).  In future work we shall show that that there is a geometric origin for the vanishing of the first order invariants, and that there is consequently a geometric way of identifying the horizon, not only for stationary black hole spacetimes but also for dynamical ones.

\newpage

\section*{Acknowledgements}  
 
The work was supported by NSERC of Canada.

\providecommand{\href}[2]{#2}
\begingroup\raggedright

\endgroup


\begin{thebibliography}{10}

\bibitem{HIGHER-D-REVIEW}
R.~{Emparan} and H.~S. {Reall}, ``{Black Holes in Higher Dimensions}'', {\em
  Living Rev. Rel.} {\bfseries 11} (2008) {\bf 6},
\href{http://arxiv.org/abs/0801.3471}{{\ttfamily arXiv:0801.3471v1 [hep-th]}}.


\bibitem{Thornburg2005} J. Thornburg, ``Event and Apparent Horizon Finders for 3+1 Numerical Relativity'', {\em
  Living Rev. Rel.} {\bf 10}, 3 (2007).

\bibitem{KLA1982} A. Karlhede, U. Lindström, and J. E. Aman. ``A note on a local effect at the Schwarzschild sphere'', {\em Gen. Rel.  Grav. } {\bf 14}, 569-571 (1982).
 
\bibitem{AbdelqaderLake2015} M. Abdelqader, and K. Lake, ``Invariant Characterization of the Kerr Spacetime: Locating the Horizon and Measuring the Mass and Spin of Rotating Black Holes Using Curvature Invariants'', {\em Phys. Rev. D}, {\bf 91}, 084017  (2015) [arXiv:1412.8757 [gr-qc]]. 


\bibitem{PageShoom2015} D. N. Page, and A. Shoom, ``Local Invariants Vanishing on Stationary Horizons: A Diagnostic for Locating Black Holes'', {\em Phys. Rev. Let.} {\bf{114}}, 141102 (2015) [arXiv:1501.03510 [gr-qc]].  

\bibitem{CMcN} A. Forget, ``The Cartan Algorithm in Higher Dimensions'',  Masters Thesis, Dalhousie University (2016).


\bibitem{kramer}
H.~Stephani, D.~Kramer, M.~MacCallum, C.~Hoenselaers, and E.~Herlt, {\em {Exact
  solutions of Einstein's field equations}}.
\newblock Camb. Univ. Press (2003).

\bibitem{class}
A.~Coley, R.~Milson, V.~Pravda, and A.~Pravdov\'a, ``{ Classification of the
  Weyl tensor in higher-dimensions}'', {\em Class. Quant. Grav.} {\bfseries
  21}, L35--L42 (2004) 
[arXiv:gr-qc/0401008 [gr-qc]].

\bibitem{BIVECTOR}
A.~Coley and S.~Hervik, ``{Higher dimensional bivectors and classification of
  the Weyl operator}'',
  \href{http://dx.doi.org/10.1088/0264-9381/27/1/015002}{{\em Class. Quant.
  Grav.} {\bfseries 27}, 015002 (2010) } [arXiv:0909.1160 [gr-qc]].

\bibitem{spinorclass}
  M.~Godazgar,
  ``Spinor classification of the Weyl tensor in five dimensions'',
{\em   Class. Quant. Grav.}  {\bf 27}, 245013 (2010) 
  [arXiv:1008.2955 [gr-qc]].


\bibitem{Alignment}
R.~Milson, A.~Coley, V.~Pravda, and A.~Pravdov\'a, ``{Alignment and
  algebraically special tensors in Lorentzian geometry}'',
  \href{http://dx.doi.org/10.1142/S0219887805000491}{{\em Int. J. Geom. Meth.
  Mod. Phys.} {\bfseries 2}, 41 (2005) }
[arXiv:gr-qc/0401010 [gr-qc]].





\bibitem{AlignmentReview}
A.~Coley, ``{Classification of the Weyl Tensor in Higher Dimensions and
  Applications}'', \href{http://dx.doi.org/10.1088/0264-9381/25/3/033001}{{\em
  Class. Quant. Grav.} {\bfseries 25}, 033001} (2008) 
[arXiv:0710.1598 [gr-qc]].

\bibitem{PraPraOrt07}
V.~Pravda, A.~Pravdov\'a, and M.~Ortaggio, ``{Type D Einstein spacetimes in
  higher dimensions}'',
  \href{http://dx.doi.org/10.1088/0264-9381/24/17/009}{{\em Class. Quant.
  Grav.} {\bfseries 24}, 4407 (2007) } [arXiv:0704.0435 [gr-qc]]

\bibitem{higherghp}
M.~N. Durkee, V.~Pravda, A.~Pravdov{\' a}, and H.~S. Reall, ``{Generalization
  of the Geroch-Held-Penrose formalism to higher dimensions}'',
  \href{http://dx.doi.org/10.1088/0264-9381/27/21/215010}{{\em Class. Quant.
  Grav.} {\bfseries 27},  215010 (2010)} [arXiv:1002.4826 [gr-qc]].

\bibitem{mahdi}
M.~{Godazgar} and H.~S. {Reall}, ``{Algebraically special axisymmetric
  solutions of the higher-dimensional vacuum Einstein equation}'',
  \href{http://dx.doi.org/10.1088/0264-9381/26/16/165009}{{\em Class. Quant.
  Grav.} {\bfseries 26}, 165009 (2009) } [arXiv:0904.4368 [gr-qc]].

\bibitem{inv}
A.~Coley, S.~Hervik, and N.~Pelavas, ``{Spacetimes characterized by their
  scalar curvature invariants}'',
  \href{http://dx.doi.org/10.1088/0264-9381/26/2/025013}{{\em Class. Quant.
  Grav.} {\bfseries 26}, 025013 (2009) } [arXiv:0901.0791 [gr-qc]].


\bibitem{DISCRIM}
A.~Coley and S.~Hervik, ``{Discriminating the Weyl type in higher dimensions 
using scalar curvature invariants}'',
{\em Gen. Rel. and Grav.}  {\bfseries 43}  (2011);
see also [arXiv:1011.2175[gr-qc]].



\bibitem{OP}
S.~Hervik and A.~Coley, ``{Curvature operators and scalar curvature
  invariants}'', \href{http://dx.doi.org/10.1088/0264-9381/27/9/095014}{{\em
  Class. Quant. Grav.} {\bfseries 27}, 095014 (2010) }
[arXiv:1002.0505 [gr-qc]].

\bibitem{CHDG}
A. A.~Coley, S.~Hervik, M. N.~Durkee and M.~Godazgar 
, ``Algebraic classification of five-dimensional spacetimes using scalar invariants'', {\em Class. Quant. Grav.} {\bf 28}, 155016 (2011) [arXiv:1105.2355[gr-qc]]. 



\bibitem{SBR}
H.~Elvang, R.~Emparan, D.~Mateos, and H.~S. Reall, ``{A supersymmetric black
  ring}'', \href{http://dx.doi.org/10.1103/PhysRevLett.93.211302}{{\em Phys.
  Rev. Lett.} {\bfseries 93}, 211302 (2004)} [arXiv:0407065[hep-th]].


\bibitem{BMPV}
J.~C. Breckenridge, R.~C. Myers, A.~W. Peet, and C.~Vafa, ``{D-branes and
  spinning black holes}'',
  \href{http://dx.doi.org/10.1016/S0370-2693(96)01460-8}{{\em Phys. Lett.}
  {\bfseries B391}, 93 (1997) }
[arXiv:9602065 [hep-th]].


\bibitem{GANG} D. Brooks, P. Chavy-Waddy, A. Coley, A. Forget, D. Gregoris, M. A. H. MacCallum, and D. McNutt, ``Cartan Invariants as Event Horizon Detectors'',~preprint~(2016).


\bibitem{RBR}
R.~{Emparan} and H.~S. {Reall}, ``A rotating black ring in five dimensions'',
  {\em Phys. Rev. Lett.} {\bfseries 88}, 101101 (2002) 
[arXiv:0110260 [hep-th]].

\bibitem{EE:2003}H. Elvang, R. Emparan, ``Black rings, supertubes, and a stringy resolution of black hole non-uniqueness'', {\em JHEP} {\bf 0311}, 035 (2003).

\bibitem{CH2011} M. Ortaggio, V. Pravda, and A. Pravdova, "Algebraic Classification of Higher-dimensional Spacetimes Based on Null Alignment", {\em Class. Quant. Grav. } {\bf 30}, 1 (2011).

\bibitem{CHOL2012} A. Coley, S. Hervik, M. Ortaggio, and L. Wylleman, "Refinements of the Weyl Tensor Classification in Five Dimensions", {\em Class. Quant. Grav. } {\bf 29}, 15 (2012).

\bibitem{brwands}
V.~Pravda and A.~Pravdov\'a, ``{WANDs of the black ring}'',
  \href{http://dx.doi.org/10.1007/s10714-005-0110-3}{{\em Gen. Rel. Grav.}
  {\bfseries 37}, 1277 (2005) }
[arXiv:0501003 [gr-qc]].

\bibitem{bdcond} Ortaggio, Marcello. ``Bel–Debever criteria for the classification of the Weyl tensor in higher dimensions.'' {\em Class. Quant. Grav.} {\bf 26}, 19, 195015 (2009).
%

\bibitem{CH}
A. A.~Coley, and S.~Hervik, `` Algebraic classification of spacetimes using discriminating scalar curvature invariants'', (2010) [arXiv:1011.2175 [gr-qc]]. 








\bibitem{5Dclass}
A.~Coley and N.~Pelavas, ``{Classification of higher dimensional spacetimes}'',
  \href{http://dx.doi.org/10.1007/s10714-006-0232-2}{{\em Gen. Rel. Grav.}
  {\bfseries 38} 445 (2006) }
[arXiv:0510064 [gr-qc]].








 


\bibitem{ref1}
J. M. Collins, R. A. d'Inverno, and J. A. Vickers, \lq\lq The Karlhede classification of type D vacuum spacetimes\rq\rq, {\em Class. Quant. Grav. }
7 2005 (1990) ; {\it,ibid.}, {\em Class. Quant. Grav.} 10  343-351 (1993).




\bibitem{ref5}
M. A. H. MacCallum, ``Spacetime Invariants and their uses'', (2015) [arXiv:1504.06857v1 [gr-qc]].




\end{thebibliography}
\end{document}